\documentclass[twoside,leqno,twocolumn]{article}

\usepackage{ltexpprt}
\usepackage{booktabs} 
\usepackage{enumerate}
\usepackage{amsmath}
\usepackage{amsfonts}
\usepackage{graphicx}
\usepackage{algorithm}
\usepackage{algpseudocode}
\usepackage{bm}
\usepackage{silence}
\usepackage{subfigure}
\usepackage{multirow}
\usepackage{array}
\usepackage{xcolor}
\usepackage{setspace}

\def\BibTeX{{\rm B\kern-.05em{\sc i\kern-.025em b}\kern-.08em
    T\kern-.1667em\lower.7ex\hbox{E}\kern-.125emX}}

\begin{document}

\title{\Large Session-based Recommendation with Hypergraph Attention Networks}
\author{Jianling Wang\thanks{Texas A\&M University, \{jlwang, zhuziwei, caverlee\}@tamu.edu} \and Kaize Ding\footnotemark[2]\thanks{Arizona State University, kaize.ding@asu.edu} \and Ziwei Zhu\footnotemark[1] \and James Caverlee\footnotemark[1]}
\date{}

\maketitle



\fancyfoot[R]{\scriptsize{Copyright \textcopyright\ 2021 by SIAM\\
Unauthorized reproduction of this article is prohibited}}




\begin{abstract} 
\small
\baselineskip=9pt 
{\color{black} Session-based recommender systems aim to improve recommendations in short-term sessions that can be found across many platforms. A critical challenge is to accurately model user intent with only limited evidence in these short sessions. For example, is a flower bouquet being viewed meant as part of a wedding purchase or for home decoration? Such different perspectives greatly impact what should be recommended next. Hence, this paper proposes a novel session-based recommendation system empowered by hypergraph attention networks. Three unique properties of the proposed approach are: (i) it constructs a hypergraph for each session to model the item correlations defined by various contextual windows in the session simultaneously, to uncover item meanings; (ii) it is equipped with hypergraph attention layers to generate item embeddings by flexibly aggregating the contextual information from correlated items in the session; and (iii) it aggregates the dynamic item representations for each session to infer the general purpose and current need, which is decoded to infer the next interesting item in the session. Through experiments on three benchmark datasets, we find the proposed model is effective in generating informative dynamic item embeddings and providing more accurate recommendations compared to the state-of-the-art.
}

\end{abstract}

\section{Introduction}
Recommendation systems are ubiquitous, acting as an essential component in online platforms to help users discover items of interest. In many practical scenarios, a recommendation system needs to infer the next interaction for a user based only on the short-term prior interactions within a particular session. Previous approaches that learn a static model of users \cite{he2018adversarial,he2017neural,rendle2012bpr,wang2019neural,zhu2019improving} or rely on long-term user behaviors \cite{he2017translation,li2020time,wang2019recurrent,wang2020time} are not well-suited for such scenarios and could lead to poor predictive power. In contrast, \textit{session-based recommendation} has attracted increasing attention, with recent approaches showing promising performance in inferring user interests in these (often anonymous) short-term sessions \cite{li2017neural,liu2018stamp,ren2019repeatnet,rendle2010factorizing,wu2019session}.

\begin{figure}
\centering
\includegraphics[width=0.48\textwidth]{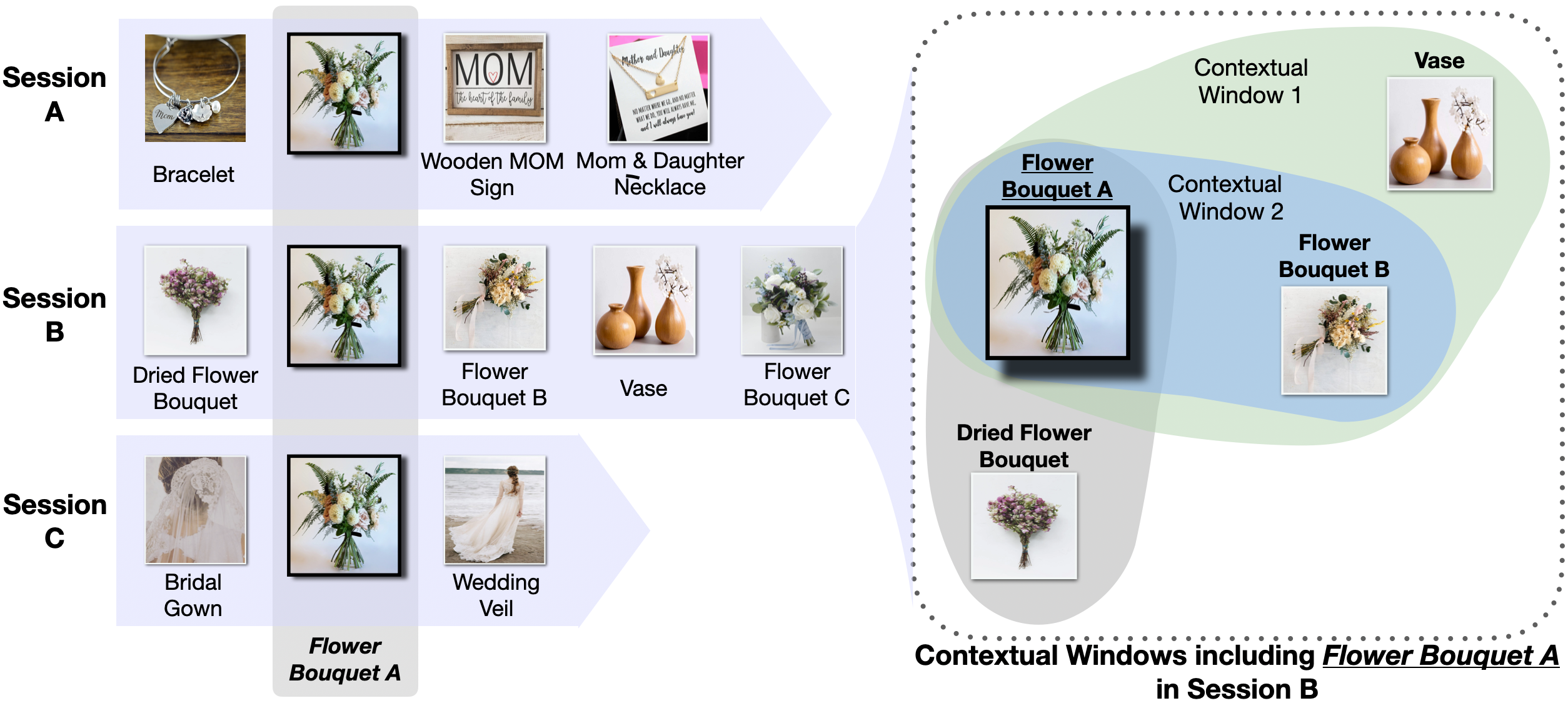}
\caption{An example with three sessions. The same flower bouquet appears in each session, but with a different purpose in each. The righthand side shows contextual windows of size 2 and 3 including this flower bouquet. Items in the same contextual window are correlated; and different contextual windows may have different levels of importance for characterizing an item.}
\label{fig:explain}
\end{figure}

A critical issue is how items are treated in such session-based recommendation approaches. The \textit{individual} items can reveal user intent, but they only provide limited evidence. To illustrate, consider the example in Figure \ref{fig:explain}. In different sessions, the same flower bouquet can be viewed differently, i.e., as part of a wedding party purchase, an option for home decoration, or a gift for Mother's Day. However, if we independently consider Flower Bouquet A, it might be viewed as exactly the same item across sessions. In a sense, the meaning of an item (and what it reveals about user intent) could be inferred from \textbf{\textit{contextual windows}}, each of which contains a set of consecutive items showing up together within a session. In Session $C$, bouquet $A$ is likely for a wedding party since the user clicks it right between other wedding-related items, while the bouquet in Session $A$ may be a Mother's Day gift since it is clicked along with items specified for ``Mom.''

Hence, we propose to exploit these contextual windows to model \textit{session-wise item representations} that can robustly capture user intent with only limited evidence available in short sessions. However, there are several key challenges in eliciting the user intent signal among items from the contextual windows in each session: \textbf{(1)} conventional graph structures and graph neural networks \cite{berg2017graph,ding2019deep,gilmer2017neural,hamilton2017inductive,kipf2016semi} are designed to model the pairwise connections between items, which are not always sufficient since we need to consider contextual windows connecting various numbers of items ranging from two to many. For example, for Session $B$ in Figure \ref{fig:explain}, the pairwise linkage between bouquet $A$ and $B$ is not enough to reveal that bouquet $A$ is meant for home decoration. But such evidence could be inferred by analyzing the triadic relations among both the bouquets and a vase as defined by Contextual Window $1$. To this end, we adopt the hypergraph \cite{bai2019hypergraph,ding2020more,feng2019hypergraph,wang2020next} structure to model the correlations amongst items in different contextual windows within each session. In a session hypergraph, each node denotes an item, and a hyperedge connects the collection of items that show up together within a contextual window. This hypergraph structure supports capturing correlations among items defined by contextual windows, which could be arbitrary-order depending on the usage scenario; \textbf{(2)} while propagating and aggregating the user intent evidence within a session hypergraph, some items are informative, but others may not be. Moreover, different contextual windows may bring in different levels of evidence for how items are represented. Thus, a key challenge is how to carefully highlight the informative items on each hyperedge and also emphasize the evidence from hyperedges with larger impacts.

To tackle the aforementioned challenges, we propose SHARE: a \underline{\textbf{S}}ession-based \underline{\textbf{H}}ypergraph \underline{\textbf{A}}ttention Network for \underline{\textbf{RE}}commendation. Specifically, the proposed SHARE has three unique characteristics:
\begin{itemize}
\setlength\itemsep{-0.15em}
\item First, it captures contextual information with sliding windows, whereby items appearing in the same window are connected with a hyperedge. By applying multiple sliding windows on the session sequence, it is able to model a session considering contextual windows of varying sizes simultaneously in the hypergraph structure.

\item Second, it incorporates a carefully-designed hypergraph attention network to extract the user intent evidence from contextual windows, which is able to pay more attention on the informative items (nodes) and also emphasize the evidence from contextual windows (hyperedges) with larger impacts.

\item Third, the session-wise item embeddings resulting from a stack of hypergraph attention layers can then be fed into a self-attention layer to infer both the general intent and current interests in the session, which are decoded jointly to generate the next-item recommendation for this session. 

\end{itemize}

With experiments on three benchmark datasets for session-based recommendation, we show that such a hypergraph-based approach is better suited than conventional graphs for modeling item correlations within sessions. The proposed SHARE is effective in predicting the next interesting items, and significantly outperforms the state-of-the-art in session-based recommendation.

\section{Related Work}

\subsubsection*{Session-based Recommendation.} Given sequential user behaviors in a session, session-based recommendation systems aim to infer the subsequent behavior.
Since these sessions usually exclude login information and take place in the short-term (e.g., less than a half hour) where user intent is local to the session, previous works that elicit user preferences from long-term historic behaviors or generate user embeddings based on their identification \cite{he2017neural,wang2019recurrent,wang2020time} are not well-suited to providing recommendations for these anonymous short-term sessions. Though matrix factorization-based and neighbor-based methods can be applied for session-based recommendation \cite{linden2003amazon,sarwar2001item}, they ignore the transitional patterns between items. Thus, FPMC \cite{rendle2010factorizing} has been proposed to extend matrix factorization with Markov Chains, which is able to predict the next item considering the transition from the last item. 
With the advancement of neural networks, different neural mechanisms have been applied to handle sequential session data. In \cite{hidasi2018recurrent,hidasi2015session}, the authors adopt recurrent neural networks to learn sequential patterns from all sessions and infer the next items with the output of the last layer. NARM \cite{li2017neural} utilizes an attention layer to aggregate items in the session and capture the main purpose of each session. Furthermore, STAMP \cite{liu2018stamp} designs another attention component to emphasize the short-term interest in the session. To capture the local dependencies between items, SR-GNN \cite{wu2019session} applies the conventional graph neural network to model the transitions between items in a session graph. In contrast, we propose to utilize hypergraphs and an attention mechanism to model item correlations and user intent with the contextual windows in session-based recommendation.

\subsubsection*{Graph-based Recommendation.}
Since graphs can be a good fit for modeling the interactions between item-item, user-user, or user-item, there are works in developing recommendation systems centering around different graph structures. 
Translation-based recommendation systems \cite{he2017translation,pasricha2018translation} treat all the items as nodes and users as the connections between items consumed subsequently. These models embed users and items into a similar space by minimizing the translation loss, which requires clear user identification (which is often not available in session-based scenarios). Recently, several efforts apply newly introduced Graph Neural Networks (GNNs) for recommendation \cite{berg2017graph,wang2019neural,zheng2018spectral}, in which different GNN models have been designed for representation learning in graph structured data. GCMC \cite{berg2017graph} consists of neural graph autoencoders to reconstruct the user-item rating graph. And in NGCF \cite{wang2019neural}, the authors propose to construct a user-item bipartite graph and utilize multiple graph neural layers to capture the multi-hop collaborative signals between users and items. In social recommendation, the relations between users can be exploited with the social graph \cite{fan2019graph,wu2019session}. For example, DiffNet \cite{fan2019graph} adopts a Graph Convolutional Network to model the diffusion of user embeddings among their social connections. Unlike SHARE, these methods are built on conventional graph structures (not hypergraphs). And they are not designed for the special characteristics of session-based recommendation, like sequential user actions in short-lived sessions without user identification.





\section{The Proposed Model - SHARE}
In this section, we propose a novel session-based recommendation model with Hypergraph Attention Networks to exploit the contextual windows within each individual session. In the following, we begin with the problem setup and then structure the design of SHARE around three research questions: (i) Given that items falling into the same contextual window are correlated, how do we construct a hypergraph for each session to model the correlations among items from a variety of contextual windows simultaneously? (ii) Considering these contextual windows, how do we update an item embedding with the user intent evidence that propagates in and across the contextual windows? (iii) With these session-wise item embeddings, how can we infer the next interesting item by extracting both the general interest and current need in the session?

\subsection{Problem Setup.}
In session-based recommendation \cite{li2017neural,ren2019repeatnet,wu2019session}, given the sequence of items which have been interacted within a session, the goal is to predict the next item. Let $\textbf{I}=\{i_1, i_2, ..., i_N\}$ denote the set of $N$ unique items in the system. These items start with a set of embeddings $\{\textbf{i}_1, \textbf{i}_2\, $..., $\textbf{i}_N\}$, each of which is a trainable embedding associated with the item ID. An anonymous session $s$ consisting of a sequence of $t$ actions (i.e., the items interacted within the session) can be denoted by $\textbf{s} = [i_{s,1}$, $i_{s,2}$,..., $i_{s,t}]$, in which the items are sorted in chronological order and $i_{s,p}$ represents the $p_{th}$ item interacted within the session. A session-based recommendation system should predict the next possible action (i.e., $i_{s,t+1}$) based on the previous $t$ actions. That is, we want to generate the preference scores for all the items in $\textbf{I}$ based on the sequence of actions in session $s$. Then the top-K preferred items can be treated as candidates for recommendation.

\subsection{Session Hypergraph Construction.}
\label{sec:graph_construction}

\begin{figure*}
\vspace{-0.2in}
\centering
\includegraphics[width=0.9\textwidth]{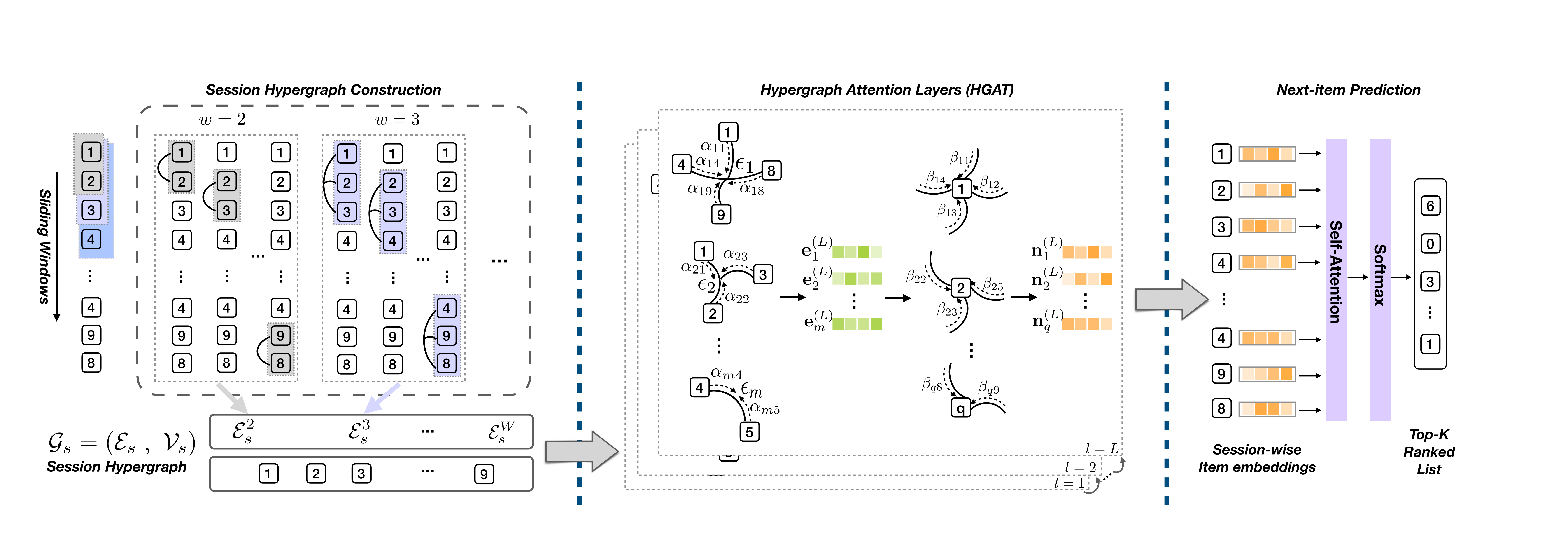}
\vspace{-0.05in}
\caption{The structure of SHARE: It applies multiple sliding windows to capture the contextual information to construct a hypergraph. With a well-designed HGAT network, it is able to generate the item embeddings revealing their meanings in the specific session. The sequence of session-wise item embeddings is fed into the self-attention layer to generate an embedding for the session, which is decoded for the preference scores on items.}
\label{fig:model}
\label{fig:graph}
\end{figure*}

With a sequence of interacted items in session $s$, we propose to model the session as a hypergraph before exploiting the item correlations in different (possibly, overlapping) contextual windows. Let $\mathcal G_s=\{\mathcal V_s, \mathcal E_s\}$ denote the hypergraph constructed from session $s$, in which the node set $\mathcal V_s$ consists of all the unique items appearing in this session. Each hyperedge $e \in \mathcal E_s$ will connect all the items falling into the specific contextual window. As in Figure \ref{fig:graph}, we can apply a sliding window with size $w$ on the item sequence for the session to identify all of the contextual windows of size $w$ in this session, with which the items appearing in the same window can be treated as items falling into the same contextual window and thus will be connected with a hyperedge. Hence, to exploit contextual windows of varying sizes in the session, we can apply sliding windows of varying sizes on the item sequence. Let $\mathcal E_s^w$ represent the collection of all the hyperedges constructed with such a sliding window of size $w$ on session $s$. Then we gather hyperedges based on different sliding windows together to be the set of hyperedges $\mathcal E_s$ for session $s$ with $\mathcal E_s = \mathcal E_s^2 \cup \mathcal E_s^3 \cup ... \mathcal E_s^W$, in which $W$ is the maximum size of contextual windows that we consider in the model. We repeat this process to construct the unique hypergraph for each of the sessions.



\subsection{Hypergraph Attention Networks.}

With the hypergraph structure, we want to learn the representations for nodes (items) considering the correlations among nodes (items) defined by different hyperedges. In the following, we introduce a novel hypergraph attention network (HGAT) which is able to aggregate the user intent evidence that propagates in and across contextual windows while updating the node representations in a hypergraph. There are recent works extending neural networks from a conventional graph to a hypergraph by generalizing the convolution operation \cite{bai2019hypergraph,feng2019hypergraph,wang2020next}, with which the neighboring node features will be firstly aggregated to the common hyperedges and then propagated to the node. They usually treat nodes \textit{equally} while characterizing the hyperedge or use a pre-defined weight matrix to model the importance of information propagated via different hyperedges. However, in session hypergraphs, for nodes belonging to the same contextual window, some of them may be informative in revealing user intent, but others may not be. And the evidence propagating via different hyperedges (contextual windows) may bring in different levels of impacts to an item. Thus, we require a new representation learning process for hypergraphs to capture these special needs. 

To tackle this problem, we propose to generalize the attention mechanism for hypergraphs so that we can \textit{highlight the informative items on each hyperedge and also emphasize the evidence from hyperedges with larger impacts}. In the following, we will explain the node representation learning process with the proposed hypergraph attention layer (see Figure \ref{fig:graph}) in two steps: 


\smallskip
\noindent\textbf{Node to Hyperedge.} Since information can propagate among neighboring nodes via hyperedges, they are the key factor for node representation learning in hypergraph. With the special hypergraph structure, instead of directly updating each node with the the neighboring node information, we firstly treat each hyperedge as an interlayer between nodes and aggregate all the information propagating via the hyperedge.


We take the graph $\mathcal{G}_s$ constructed by session $s$ as an example. The operation can be applied on hypergraphs for other sessions. Let $\{\textbf{n}_1^{(0)}, \textbf{n}_2^{(0)},..., \textbf{n}_{p}^{(0)}\} = \{\textbf{i}_1, \textbf{i}_2,..., \textbf{i}_{p}\}$ denote the node input for the first HGAT layer, which is the initial item embeddings for the set of unique items $\{i_1, i_2,..., i_{p}\}$ appearing in $\mathcal{G}_s$. 
Since some nodes on a hyperedge are informative but others may not be, we should pay varying attention on the information from these nodes while aggregating them together. Let $\textbf{m}_{t \sim j}^{(1)}$ denote the information propagating via hyperedge $\epsilon_j$ from node $t$ on the $(1)_{th}$ HGAT layer. We can aggregate the information from each of the nodes connected by $\epsilon_j$ with the attention operation to generate the representation $\textbf{e}_j^{(1)}$ as:
\begin{equation}
\begin{aligned}
\textbf{e}_j^{(1)} &= \sum_{t \in \mathcal{N}_j} \textbf{m}_{t \sim j}^{(1)} \quad\quad \textbf{m}_{t \sim j}^{(1)} = \alpha_{jt} \textbf{W}_1^{(1)} \textbf{n}_t^{(0)}\\
\alpha_{jt} &= \frac{S(\hat{\textbf{W}}_1^{(1)}\textbf{n}_t^{(0)}, \textbf{u}^{(1)})}{\sum_{f \in \mathcal{N}_j} S( \hat{\textbf{W}}_1^{(1)}\textbf{n}_f^{(0)}, \textbf{u}^{(1)})}
\end{aligned}
\end{equation}
in which $\mathcal{N}_j$ denotes all the nodes connected by hyperedge $\epsilon_j$ and $\textbf{u}^{(1)}$ represents a trainable node-level context vector for the $(1)_{th}$ HGAT layer. $\textbf{W}_1^{(1)}$ and $\hat{\textbf{W}}_1^{(1)}$ are the transform matrices and $\alpha_{jt}$ denotes the attention score of node t on hyperedge $\epsilon_j$. We use a function $S(\cdot,\cdot)$ to calculate the similarity between the node embedding and context vector. Empirically, we use \textit{Scaled Dot-Product Attention} to calculate the attention scores \cite{kang2018self,vaswani2017attention}, which is defined as:
\begin{equation}
S(\textbf{a}, \textbf{b}) = \frac{\textbf{a}^T\textbf{b}}{\sqrt{D}}
\label{equ:scaled-dot}
\end{equation}
where $D$ is the dimension size and can be used for normalization while calculating the similarity scores.


\smallskip
\noindent\textbf{Hyperedge to Node.} To update the embedding for a node, we need to aggregate the contents from all its connected hyperedges. Similarly, we utilize the attention mechanism to perform the aggregation in order to model the significance of different hyperedges. Let $\textbf{m}_{j \rightarrow t}^{(1)}$ denote information (user intent evidence) from hyperedge $\epsilon_j$ to node $t$. Given the set of hyperedges $\mathcal{Y}_t$ which are connected to node $t$, its update embedding is calculated as:
\begin{equation}
\begin{aligned}
\textbf{n}_t^{(1)} &= \sum_{j \in \mathcal{Y}_t} \textbf{m}_{j \rightarrow t}^{(1)} \quad\quad \textbf{m}_{j \rightarrow t}^{(1)} = \beta_{tj} \textbf{W}_2^{(1)} \textbf{e}_j^{(1)} \\
\beta_{tj} &= \frac{S(\hat{\textbf{W}}_2\textbf{e}_j^{(1)}, \textbf{W}_3^{(1)}\textbf{n}_t^{(0)})}{\sum_{f \in \mathcal{Y}_t} S(\hat{\textbf{W}}_2^{(1)}\textbf{e}_f^{(1)}, \textbf{W}_3^{(1)}\textbf{n}_t^{(0)})}
\end{aligned}
\end{equation}
where $\textbf{W}_2^{(1)}$, $\hat{\textbf{W}}_2^{(1)}$ and $\textbf{W}_3^{(1)}$ are the trainable matrices used to transform the vector before calculating the attention scores for the $(1)_{th}$ HGAT layer. $\beta_{tj}$ indicates the impact of hyperedge $\epsilon_j$ on node $t$. As in the last step, we use the Scaled Dot-Product Attention formula defined by Equation \ref{equ:scaled-dot} to calculate $S$. The resulting $\textbf{n}_t^{(1)}$ can be treated as the updated embedding for note $t$ by aggregating information from its neighboring region in the hypergraph.


\smallskip
\noindent\textbf{High-order Propagation.} While a single HGAT layer can capture the information from direct neighbors, we construct a Hypergraph Attention Network by stacking multiple HGAT layers to model the multi-hop high-order information propagation in the hypergraph. In this Hypergraph Attention Network, the output of the $(l-1)^{th}$ HGAT layer is the input for the $l^{th}$ layer. Thus the outputs (i.e., node embeddings) from the last layer ($L$) inherit the contextual information from all the previous layers and can be used to characterize the items in this session. For each of the session hypergraph, such an Hypergraph Attention Network is able to generate session-wise item embeddings to reflect user intent in the session, by highlighting the informative items on each hyperedge and also emphasizing the evidence from hyperedges with larger impacts.


\subsection{Next-Item Prediction.}
To infer the next interesting item, our goal is to generate an embedding which can encode both the general interest and the current need of the session. We adopt the idea from self-attention \cite{kang2018self,vaswani2017attention} to achieve this goal. Since the general interest can be uncovered by the aggregation of all the items and the current need is revealed by the last item, we want to aggregate the items in the session while paying more attention on those items which are highly correlated to the last one. Thus, we can \textit{treat the last item in the session as the query and the sequence of items in the session as both keys and values}, leading to the design of the self-attention layer in Figure \ref{fig:model}. 





For a session interacted with $[i_{s,1}, i_{s,2}, ...,i_{s,t}]$ sequentially, we will lookup the corresponding node embeddings from the output of the hypergraph attention networks to get $[\textbf{n}_{s,1}^{(L)}, \textbf{n}_{s,2}^{(L)}, ..., \textbf{n}_{s,t}^{(L)}]$. We will transform the item embeddings with $\textbf{W}_Q$, $\textbf{W}_K$ and $\textbf{W}_V$ to generate the query vectors, key vectors, and value vectors correspondingly. Then we can aggregate the sequence of embeddings with:
\begin{equation}
\begin{aligned}
\textbf{h}_s &= \sum_{i \leq t} \sigma_{ti} \textbf{W}_V\textbf{n}_{s,i}^{(L)} \\
\sigma_{ti} &= \frac{S(\textbf{W}_Q\textbf{n}_{s,t}^{(L)}, \textbf{W}_K\textbf{n}_{s,i}^{(L)})}{\sum_{j \leq t} S(\textbf{W}_Q\textbf{n}_{s,t}^{(L)}, \textbf{W}_K\textbf{n}_{s,j}^{(L)})}
\end{aligned}
\end{equation}
in which the attention scores $S$ is defined in Equation \ref{equ:scaled-dot}. According to observations in previous research \cite{kang2018self,wu2019session}, the order of items is less likely to be related to the general interest in short-term sequences and could introduce noise in modeling such short-term sequences. We omit the order information while handling the item sequences in session-based recommendation.

With this carefully-designed self-attention layer, the resulting $\textbf{h}_s$ encodes both the general interest and the current need for session $s$. Then we compute the multiplication between $\textbf{h}_s$ and the latent factor of item $v$ to predict the preference score of session $s$ on $v$ using $p_{sv} = \textbf{h}_s^T\textbf{i}_v$.  Let $\textbf{p}_s = [p_{s1},p_{s2},...,p_{sN}]$ denote the predicted preference scores of session $s$ on all of the $N$ items in the system, which will be processed with a Softmax layer to generate the final scores such that: $\hat{\textbf{y}}_s = softmax(\textbf{p}_s)$. We use a one-hot vector $\textbf{y}_s = [y_{s1},y_{s2},...,y_{sN}]$ to denote the ground-truth item of session $s$. During the training phrase, we calculate the cross-entropy loss for all training sessions $S_{train}$ with $\mathcal{L} = - \sum_{s \in S_{train}} \sum_{v=1}^{N} y_{sv}\log{\hat{y}_{sv}}$.
Then we can train the model using back-propagation. In the testing phase, given an unseen session, we can construct a new hypergraph and feed it into the hypergraph attention network. And then we can calculate its preference scores on all the items and recommend the items ranked among the top. 

\section{Experiments}
In this section, we conduct several experiments to evaluate how the proposed SHARE model performs in session-based recommendation. 

\begin{table}[t]
\centering
\scalebox{0.78}{
\begin{tabular}{c|cccc}
\hline
 & \begin{tabular}[c]{@{}c@{}}\# Sessions\\ (Training)\end{tabular} & \begin{tabular}[c]{@{}c@{}}\# Sessions\\ (Testing)\end{tabular} & \# Items & \begin{tabular}[c]{@{}c@{}}Avg. Length \\ of Sessions\end{tabular} \\ \hline
\textbf{\textbf{$YooChoose_{1/64}$}} & 369,859 & 55,898 & 16,766 & 6.16 \\ \hline
\textbf{\textbf{$YooChoose_{1/4}$}} & 5,917,745 & 55,898 & 29,618 & 5.71 \\ \hline
\textbf{$Diginetica$} & 719,470 & 60,858 & 43,097 & 5.12 \\ \hline
\end{tabular}%
}
\caption{Dataset Statistics.}
\label{tab:data}
\end{table}

\subsection{Data.}

We adopt two public datasets that have been widely used to evaluate session-based recommendation: YooChoose and Diginetica 
\cite{hidasi2015session,li2017neural,ren2019repeatnet,wu2019session}. Yoochoose contains sessions of click events from an online retailer in Europe and was originally published as part of the 2015 RecSys Challenge. Diginetica contains sessions of product transaction data from an online retailer and was released as part of the 2016 CIKM Cup.

\begin{table*}[t]
\centering
    \resizebox{\textwidth}{!}{
\begin{tabular}{c|c|ccccccccc|c}
\hline
\textbf{Datasets} & & \textbf{S-Pop} & \textbf{MF} & \textbf{FPMC} & \textbf{KNN} & \textbf{GRU4Rec+} & \textbf{NARM} & \textbf{STAMP} & \textbf{RPNet} & \textbf{SR-GNN} & \textbf{SHARE} \\ \hline
\multirow{2}{*}{\textbf{$YooChoose_{1/64}$}} & Hit@20 & 30.44 & 31.31 & 45.62 & 51.60 & 67.84 & 68.32 & 68.74 & 69.13 & 70.57$\dagger$ & \textbf{71.51}$\ast$ \\
 & MRR@20 & 18.35 & 12.08 & 15.01 & 21.81 & 29.00 & 28.63 & 29.67 & 30.24 & 30.94$\dagger$ & \textbf{31.45}$\ast$ \\ \hline
\multirow{2}{*}{\textbf{$YooChoose_{1/4}$}} & Hit@20 & 27.08 & 3.40 & 51.86 & 52.31 & 69.11 & 69.73 & 70.44 & 70.71 & 71.36$\dagger$ & \textbf{72.25}$\ast$ \\
 & MRR@20 & 17.75 & 1.57 & 17.50 & 21.70 & 29.22 & 29.23 & 30.00 & 31.03 & 31.89$\dagger$ & \textbf{32.11}$\ast$ \\ \hline
\multirow{2}{*}{\textbf{$Diginetica$}} & Hit@20 & 21.06 & 5.24 & 26.53 & 35.75 & 46.16 & 49.70 & 45.64 & 47.79 & 50.73$\dagger$ & \textbf{52.73}$\ast$ \\
 & MRR@20 & 13.68 & 1.98 & 6.95 & 11.57 & 14.69 & 16.17 & 14.32 & 17.66$\dagger$ & 17.59 & \textbf{18.05}$\ast$ \\ \hline
\end{tabular}}%

\caption{Comparison of Different Models. All the results are in percentage (\%). The best performing method in each column is boldfaced, and the second best method is marked with $\dagger$. $\ast$ indicates that the improvement of the best result is statistically significant compared with the next-best result with $p<0.05$.}
\label{tab:result}
\end{table*}

We keep sessions with length longer than 1 and items which appear in at least 5 sessions. For Yoochoose, we test on all the sessions happening on the last day in the dataset while training all the sessions before that. Only items appearing in the training set are considered. As for Diginetica, we split the dataset and use the sessions happening in the last 7-days for testing. Furthermore, we adopt a standard sequence preprocessing method used in previous work  \cite{liu2018stamp,tan2016improved,wu2019session} to generate the session sequences and labels. Then we get the final Diginetica dataset as described in Table \ref{tab:data}. However, it is not necessary to train on the entire set of training sequences from YooChoose since it is extremely large and training only on a fraction of recent sessions can lead to better prediction performance based on the experimental results in \cite{tan2016improved}. As in \cite{li2017neural,liu2018stamp,wu2019session}, we sort all of the training sequences generated from YooChoose, and retrieve the most recent 1/64 and 1/4 to be the training samples in $YooChoose_{1/64}$ and $YooChoose_{1/4}$ (listed in Table \ref{tab:data}). Note that $YooChoose_{1/64}$ and $YooChoose_{1/4}$ share the same set of testing samples. In addition, for fair comparison, the training samples and testing samples in all of the three datasets are exactly the same as in \cite{li2017neural,liu2018stamp,ren2019repeatnet,wu2019session}.


\subsection{Experimental Setup}
\subsubsection{Evaluation Metrics.} We aim to evaluate how each model performs in predicting the next item in each session. Thus, the experiments follow the leave-one-out setting with one ground-truth item to be tested for each session in the test set. As in previous works for session-based recommendation \cite{li2017neural,ren2019repeatnet,wu2019session}, we adopt both Mean Reciprocal Rank (MRR@K) and Hit Rate (Hit@K) as evaluation metrics. Given the ranked list of items predicted for each session, Hit@K measures the probability that the ground-truth item is within the top-K. Let $r_s$ denote the ranking of the ground-truth item for session $s$. Then, $Hit_s@K = 1$ if $r_s \le K$ and $Hit_s@K = 0$ otherwise. As for MRR@K, it measures the average ranking of the ground-truth items among the lists. That is $MRR_s@K = \frac{1}{r_s}$ if $r_s \le K$ otherwise $MRR_s@K = 0$. 
Then we take the average values of MRR and Hit Rate over all the sessions in the test set and report the results.

\begin{table*}[t]
\centering
\resizebox{0.99\textwidth}{!}{%
\begin{tabular}{l|cc|cc|cc|cc|cccc}
\hline
\multirow{3}{*}{\textbf{Methods}} & \multicolumn{4}{c|}{\textbf{$YooChoose_{1/64}$}} & \multicolumn{4}{c|}{\textbf{$YooChoose_{1/4}$}} & \multicolumn{4}{c}{\textbf{$Diginetica$}} \\ \cline{2-13} 
 & \multicolumn{2}{c|}{Hit} & \multicolumn{2}{c|}{MRR} & \multicolumn{2}{c|}{Hit} & \multicolumn{2}{c|}{MRR} & \multicolumn{2}{c|}{Hit} & \multicolumn{2}{c}{MRR} \\ \cline{2-13} 
 & @10 & @20 & @10 & @20 & @10 & @20 & @10 & @20 & @10 & \multicolumn{1}{c|}{@20} & @10 & @20 \\ \hline
(1) \textbf{w/o Hypergraph} & 59.79 & 70.45 & 29.46 & 30.21 & 60.13 & 70.79 & 29.77 & 30.52 & 37.57 & \multicolumn{1}{c|}{50.52} & 16.17 & 17.06 \\
(2) \textbf{+ GAT} & 60.25 & 70.88 & 29.77 & 30.59 & 60.60 & 71.22 & 30.54 & 31.29 & 38.36 & \multicolumn{1}{c|}{51.54} & 17.75 & 17.61 \\
(3) \textbf{+ HyperGCN} & 60.93 & 71.29 & 30.65 & 31.37 & 60.98 & 71.73 & 30.58 & 31.33 & 39.32 &  \multicolumn{1}{c|}{52.39} & 16.83 & 17.73 \\
 \hline \textbf{SHARE} & \textbf{61.13}$\ast$ & \textbf{71.51}$\ast$ & \textbf{30.67}$\ast$ & \textbf{31.45}$\ast$ & \textbf{61.52}$\ast$ & \textbf{72.25}$\ast$ & \textbf{30.96}$\ast$ & \textbf{32.11}$\ast$ & \textbf{39.52}$\ast$ & \multicolumn{1}{c|}{\textbf{52.73}$\ast$} & \textbf{17.12}$\ast$ & \textbf{18.05}$\ast$ \\ \hline
\end{tabular}%
}
\caption{Ablation Test Result. All the results are in percentage (\%). $\ast$ indicates that the improvement of the best result is statistically significant compared with the next-best result with $p<0.05$.}
\label{tab:ablation}
\end{table*}

\subsubsection{Baselines.}
\begin{itemize}
\setlength\itemsep{-0.15em}
\item \textbf{S-Pop}: This simple baseline recommends the most popular items based on their popularity in the current session. Ties are broken up using the popularity values based on the whole training set. 
\item \textbf{KNN} \cite{sarwar2001item}: This method recommends items which are most similar to items clicked in the current session. Each item is represented with a binary vector indicating all the sessions it appears in and cosine similarity is used to define their similarity. 
\item \textbf{MF} \cite{rendle2012bpr}: This matrix factorization-based model is trained with Bayesian personalized ranking loss.
\item \textbf{FPMC} \cite{rendle2010factorizing}:   Building on top of both Matrix Factorization and Markov Chains, it is able to infer the next item based on sequential behaviors. 
\item \textbf{GRU4Rec+} \cite{hidasi2018recurrent}:   This baseline utilizes Gated Recurrent Unit (GRU) to model sequential patterns, with a new ranking loss functions tailored to RNNs in session-based recommendation. 
\item \textbf{NARM} \cite{li2017neural}:  Besides learning the sequential behavior with GRU, it also includes an attention layer to extract the session's main purpose, which are combined together to infer the preference scores. 
\item \textbf{STAMP} \cite{liu2018stamp}:  This model combines both general interests from long-term memory and current interests from the short-term memory of the last-clicks in session-based recommendation. 
\item \textbf{RPNet} \cite{ren2019repeatnet}: Considering the repeat consumption in each session, this model can recommend unclicked items in the explore mode while recommending repeated items in the repeat mode. 
\item \textbf{SR-GNN} \cite{wu2019session}:  This is the state-of-the-art for session-based recommendation with graph neural networks. It models session sequences as graph-structured data and designs a graph neural network to capture the transition between items. Recommendation is made with the composition of the whole session and the last click. 
\end{itemize}
\subsubsection{Parameter Tuning.} All of the experiments are conducted on a server machine equipped with a 12 GB Titan Xp GPU. We use Adam as the optimizer with a learning rate set to be 0.001 for all datasets. The batch size for $YooChoose_{1/64}$ is 100. Since there are more training sessions in both $YooChoose_{1/4}$ and Diginetica, we set their batch size to be 300. For a fair comparison, the dimension of the item embedding is set to be 100 as in the baseline methods. We grid search for the dropout rate in \{0.1, 0.2, 0.3, 0.4, 0.5, 0.6\} and L2 regularization in \{$10^{-5}$, $10^{-4}$, $10^{-3}$, $10^{-2}$, $10^{-1}$\}. The dropout is set to be 0.3 and the L2 regularization is set to be $10^{-6}$ for all the datasets. We train each model for 50 epochs or until its performance does not improve for the validation set after 5 epochs. We fine-tune the maximum size of the sliding window in \{2, 3, 4, 5, 6\} and the number of HGAT layer in \{1, 2, 3, 4\} for different datasets. As in previous works, we randomly sample 10\% of the training sessions as validation for parameter tuning.


\subsection{Overall Evaluation.}


To compare SHARE with the baseline models, we summarize the overall results in Table \ref{tab:result}. As in previous work  \cite{li2017neural,liu2018stamp,wu2019session}, we report MRR and  Hit Rate at K=20 for the overall comparison with the baselines. We find that SHARE outperforms all of the baselines under each of the metrics for session-based recommendation over all of the datasets.


Starting from the simplest method, since MF is unable to capture sequential information, it performs poorly in session-based recommendation. Integrating with Markov Chains, FPMC can improve MF, but it is still not a good fit for modeling the short-term user behaviors and performs worse than KNN. With advances in neural networks, we observe that GRU4Rec+ achieves much better results than the non-neural models in sequential pattern modeling. However, in comparison to GRU4Rec+, we see that NARM and STAMP are better suited to extract the general interests from each short-term anonymous session by aggregating items with weighted attention scores. Emphasizing the item clicked most recently helps STAMP achieve better performance than NARM. Furthermore, by predicting and modeling the repeat consumption in sessions, RPNet can outperform the attention-based and GRU-based models in this recommendation scenario.

As the state-of-the-art for session-based recommendation, SR-GNN outperforms RPNet, verifying the effectiveness of modeling item transitions with appropriate graph structure to obtain accurate user representations in sessions. However, the conventional graph structure is insufficient in capturing the correlations amongst items across different contextual windows. Since SHARE is specifically designed with this challenge in mind, we observe that it achieves significant improvements compared with SR-GNN. Further, we observe that the performance of SHARE on $YooChoose_{1/64}$ is even better than SR-GNN on $YooChoose_{1/4}$. Though $YooChoose_{1/64}$ and $YooChoose_{1/4}$ contain different numbers of training sessions, they are tested on exactly the same set of sessions. That is, with much less training data, SHARE can recommend more accurately than the state-of-the-art model trained with more data.

\begin{figure}
\centering
\includegraphics[width=0.48\textwidth]{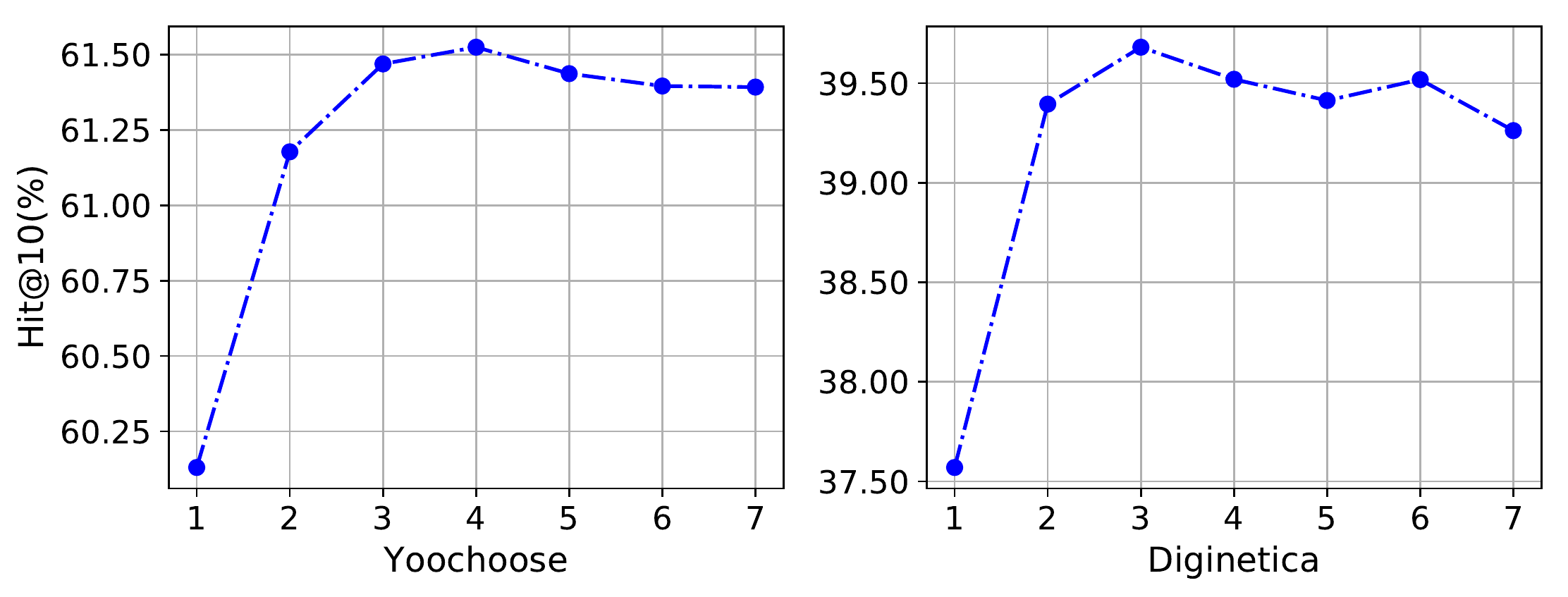}
\vspace{-0.1in}
\caption{Size of contextual windows vs. Hit Rate.}
\vspace{-0.1in}
\label{fig:window}
\end{figure}

\subsection{Ablation Analysis.}
To examine the design of the proposed model, we conduct an ablation test to compare SHARE with three of its variants and report the results in Table \ref{tab:ablation}. In Model (1), we remove the Hypergraph attention network from SHARE, meaning that a set of static item embeddings are directly fed into the self-attention layer for different sessions. Instead of utilizing the hypergraph structure, Model (2) constructs a conventional graph for each session based on \textit{pairwise} item co-occurrence within the session and use a Graph Attention Networks (GAT) \cite{velivckovic2017graph} to generate the dynamic item embeddings. In Model (3), we train a hypergraph convolution networks \cite{feng2019hypergraph} instead of HGAT network to generate the dynamic item embeddings. To compare them under different conditions, we report their performance under both $K = 10$ and $K = 20$. Overall, SHARE can outperform all of its variants for both K values, indicating the effectiveness of its design.

Without the hypergraph component, though it includes the self-attention layer to learn the sequential patterns, Model (1) achieves the weakest results since it does not consider the item correlations within each session and uses the same embedding for an item across different sessions. Then in Model (2), building on top of Model (1), we adopt the conventional graph structure and GAT networks \cite{velivckovic2017graph} to obtain dynamic item representations considering the consecutive transitions of items in each session. This model can outperform model (1), illustrating the necessity of modeling the in-session item correlations with a graph-based structure. Next, to take advantage of the hypergraph structure for modeling the complex item correlations in each session, we construct a hypergraph for each session following the same procedure as in SHARE. But in Model (3), we generate the item embedding with hypergraph convolutional networks \cite{feng2019hypergraph}, which aggregates information from the neighboring hyperedges with the convolutional operation. It can achieve better performance than using conventional graphs, which shows the effectiveness of the hypergraph structure and the necessity of modeling the contextual windows for item representation learning. However, it still falls behind SHARE since it does not consider the informativeness of nodes and the different impacts from contextual windows.


\begin{figure}
\centering
\includegraphics[width=0.48\textwidth]{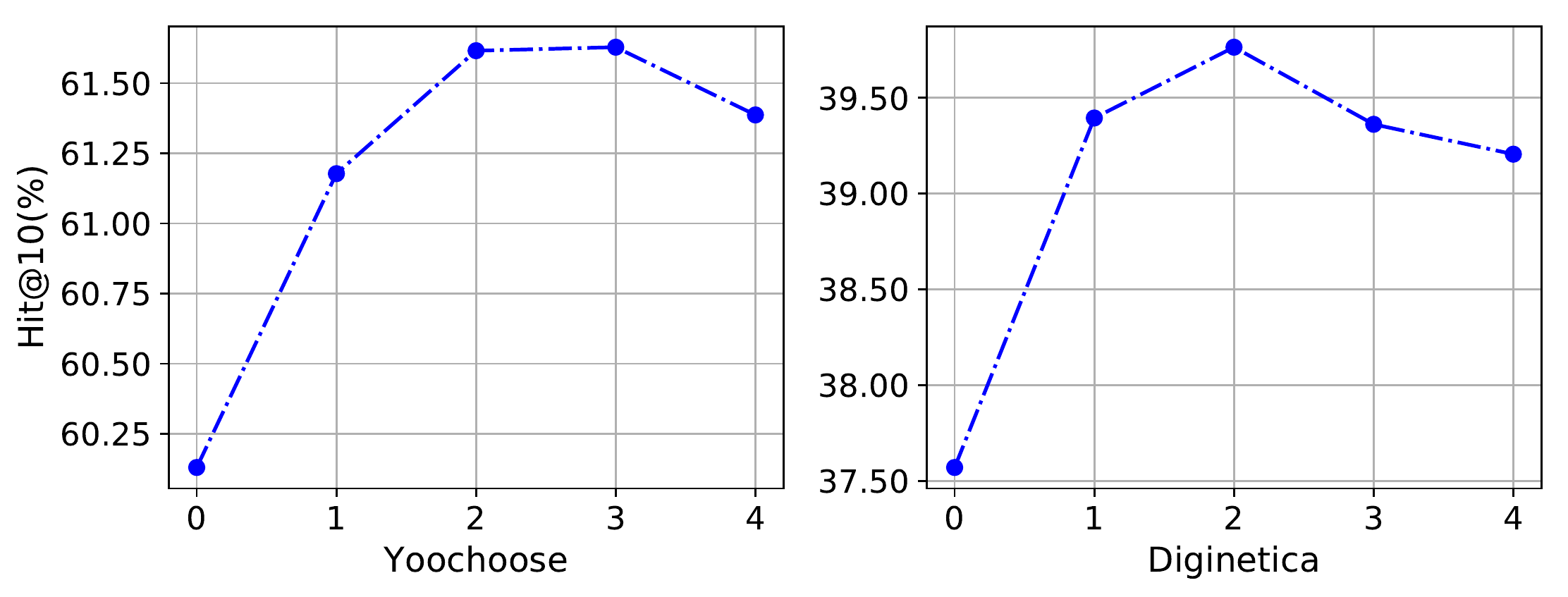}
\vspace{-0.1in}
\caption{Number of HGAT Layers vs. Hit Rate.}
\vspace{-0.1in}
\label{fig:layer}
\end{figure}

\subsection{Study of SHARE.}
In this section, we conduct further studies on three core factors in SHARE: size of the contextual window, depth of information propagation and length of the sessions. In the following, due to the limit of space, as $YooChoose_{1/64}$ and $YooChoose_{1/4}$ shares the same set of testing sessions, we show the results trained with $YooChoose_{1/4}$ to uncover the patterns for YooChoose. 

\smallskip
\noindent{\textbf{Size of contextual window.}} We first examine the performance of SHARE by varying the maximum size of contextual windows that we consider while constructing the hypergraphs (in Section \ref{sec:graph_construction}).  We use the 1-layer HGAT network to avoid the influence of high-order information propagation. Note that maximum size = 1 is similar to the case that no hypergraph is constructed for each session, meaning that items use static embeddings across sessions. As in Figure \ref{fig:window}, for YooChoose, at the beginning, the performance of SHARE improves as contextual windows with larger size are incorporated into the hypergraphs. In Diginetica, the best choice of window size is smaller that in YooChoose. The reason could be that the length of sessions is shorter in Diginetica, and thus we need only consider smaller contextual windows. These observations show that contextual windows can have significant contribution on characterizing items in each session and the proposed framework is effective in modeling various contextual windows simultaneously. 



\smallskip
\noindent\textbf{High-order Propagation.} We can stack multiple HGAT layers to model the information flowing among items with high-order connections in hypergraph. To visualize the high-order information propagation, while setting the maximum window size to be constant (i.e., 2), we show the resulting Hit Rate in Figure \ref{fig:layer} by varying the number of HGAT layers in SHARE. In each dataset, starting with a single HGAT layer, we find that the performance is improved by stacking one more layer, indicating the importance of modeling the information via high-order connections. For Diginetica, stacking more than two layers will worsen the performance since the sessions in this dataset are short in general. Using too many layers will bring in noise for the representation learning process. Meanwhile, the optimized number of layers for YooChoose is larger than that for Diginetica. Thus, we conclude that SHARE can capture both the direct and high-order connections among items in the hypergraphs and lead to accurate recommendation. 



\begin{figure}
\centering
\includegraphics[width=0.48\textwidth]{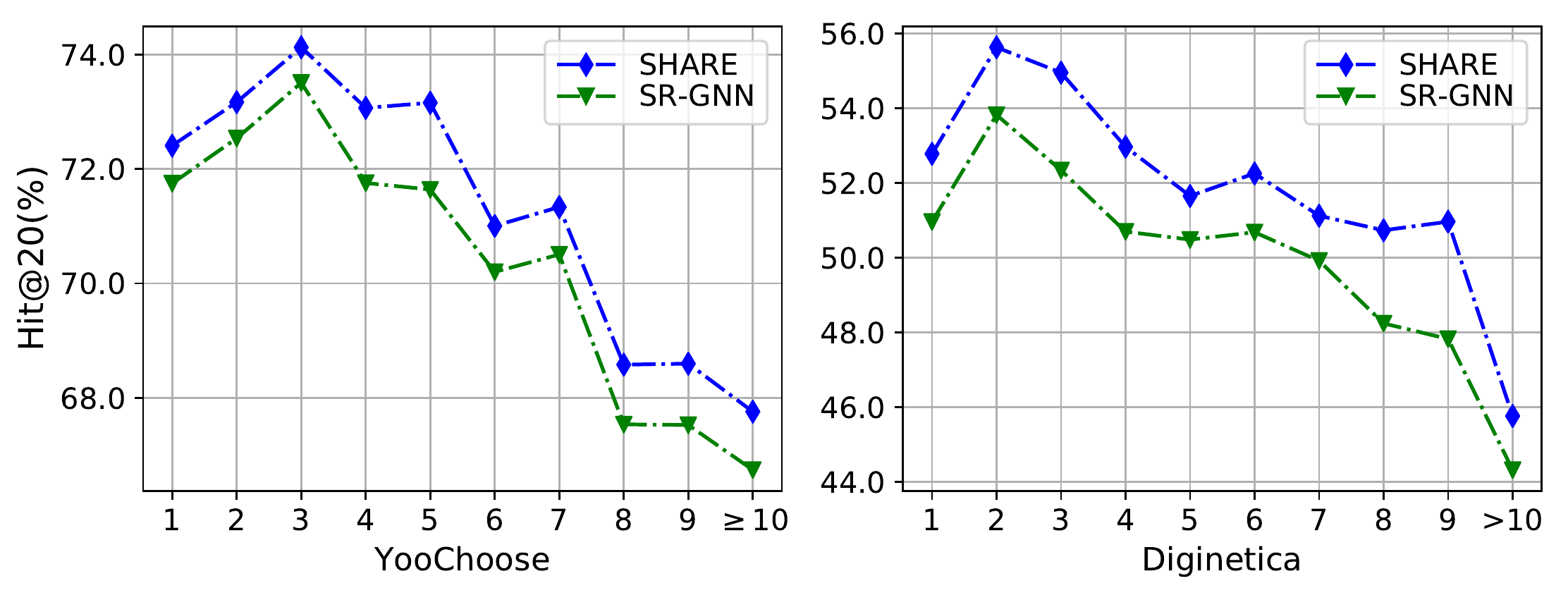}
\vspace{-0.1in}
\caption{Session Length vs. Recommendation Quality.}
\vspace{-0.1in}
\label{fig:length}
\end{figure}

\smallskip
\noindent{\textbf{Short vs. Long Sessions.}} To fully understand how SHARE performs in modeling sessions under different circumstances, we group the sessions based on their length and test how SHARE and SR-GNN perform for each group of sessions. As in Figure \ref{fig:length}, SHARE provides better recommendations for all session lengths in comparison with SR-GNN. Both models can make more accurate predictions for short action sequences than the longer ones, since they omit the order information which may be important for long sessions with adequate user behaviors. However, since the percentage of long sessions is low in real-world datasets for session-based recommendation, the main purpose is to boost the recommendation for sessions with fewer items. Indeed, the proposed framework is suitable for such practical scenarios (i.e., short-term sessions) characterized by only a limited number of sequential actions.


\section{Conclusion and Future Work}
To recommend the next interesting items in short-term sessions, we are motivated to exploit the correlations among items within various contextual windows in each session to better model their dynamic meanings across sessions. In this work, we propose a novel session-based recommendation system -- SHARE --  which is empowered by the hypergraph structure and hypergraph attention networks. 
With experiments on three real-world benchmark datasets, we find that the proposed SHARE is able to generate effective session-wise item embeddings and thus provide more accurate recommendation compared with baseline models. While in this work we mainly studying how to generate session-wise item representations, in the future we want to explore how to capture the item dynamics along time and over different domains in session-based recommendation. Meanwhile, we are also interested in extending the proposed hypergraph attention networks for representation learning in other research areas. 





\bibliographystyle{siam}

\bibliography{acmart.bib}

\begin{thebibliography}{10}

\bibitem{bai2019hypergraph}
{\sc S.~Bai, F.~Zhang, and P.~H. Torr}, {\em Hypergraph convolution and
  hypergraph attention}, arXiv preprint arXiv:1901.08150,  (2019).

\bibitem{berg2017graph}
{\sc R.~v.~d. Berg, T.~N. Kipf, and M.~Welling}, {\em Graph convolutional
  matrix completion}, arXiv preprint arXiv:1706.02263,  (2017).

\bibitem{ding2019deep}
{\sc K.~Ding, J.~Li, R.~Bhanushali, and H.~Liu}, {\em Deep anomaly detection on
  attributed networks}, in SDM, 2019.

\bibitem{ding2020more}
{\sc K.~Ding, J.~Wang, J.~Li, D.~Li, and H.~Liu}, {\em Be more with less:
  Hypergraph attention networks for inductive text classification}, in EMNLP,
  2020.

\bibitem{fan2019graph}
{\sc W.~Fan, Y.~Ma, Q.~Li, Y.~He, E.~Zhao, J.~Tang, and D.~Yin}, {\em Graph
  neural networks for social recommendation}, in WWW, 2019.

\bibitem{feng2019hypergraph}
{\sc Y.~Feng, H.~You, Z.~Zhang, R.~Ji, and Y.~Gao}, {\em Hypergraph neural
  networks}, in AAAI, 2019.

\bibitem{gilmer2017neural}
{\sc J.~Gilmer, S.~S. Schoenholz, P.~F. Riley, O.~Vinyals, and G.~E. Dahl},
  {\em Neural message passing for quantum chemistry}, in ICML, 2017.

\bibitem{hamilton2017inductive}
{\sc W.~Hamilton, Z.~Ying, and J.~Leskovec}, {\em Inductive representation
  learning on large graphs}, in NeurIPS, 2017.

\bibitem{he2017translation}
{\sc R.~He, W.-C. Kang, and J.~McAuley}, {\em Translation-based
  recommendation}, in RecSys, 2017.

\bibitem{he2018adversarial}
{\sc X.~He, Z.~He, X.~Du, and T.-S. Chua}, {\em Adversarial personalized
  ranking for recommendation}, in SIGIR, 2018.

\bibitem{he2017neural}
{\sc X.~He, L.~Liao, H.~Zhang, L.~Nie, X.~Hu, and T.-S. Chua}, {\em Neural
  collaborative filtering}, in WWW, 2017.

\bibitem{hidasi2018recurrent}
{\sc B.~Hidasi and A.~Karatzoglou}, {\em Recurrent neural networks with top-k
  gains for session-based recommendations}, in CIKM, 2018.

\bibitem{hidasi2015session}
{\sc B.~Hidasi, A.~Karatzoglou, L.~Baltrunas, and D.~Tikk}, {\em Session-based
  recommendations with recurrent neural networks}, in ICLR, 2016.

\bibitem{kang2018self}
{\sc W.-C. Kang and J.~McAuley}, {\em Self-attentive sequential
  recommendation}, in ICDM, 2018.

\bibitem{kipf2016semi}
{\sc T.~N. Kipf and M.~Welling}, {\em Semi-supervised classification with graph
  convolutional networks}, arXiv preprint arXiv:1609.02907,  (2016).

\bibitem{li2017neural}
{\sc J.~Li, P.~Ren, Z.~Chen, Z.~Ren, T.~Lian, and J.~Ma}, {\em Neural attentive
  session-based recommendation}, in CIKM, 2017.

\bibitem{li2020time}
{\sc J.~Li, Y.~Wang, and J.~McAuley}, {\em Time interval aware self-attention
  for sequential recommendation}, in WSDM, 2020.

\bibitem{linden2003amazon}
{\sc G.~Linden, B.~Smith, and J.~York}, {\em Amazon. com recommendations:
  Item-to-item collaborative filtering}, IEEE Internet computing, 7 (2003),
  pp.~76--80.

\bibitem{liu2018stamp}
{\sc Q.~Liu, Y.~Zeng, R.~Mokhosi, and H.~Zhang}, {\em Stamp: short-term
  attention/memory priority model for session-based recommendation}, in KDD,
  2018.

\bibitem{pasricha2018translation}
{\sc R.~Pasricha and J.~McAuley}, {\em Translation-based factorization machines
  for sequential recommendation}, in RecSys, 2018.

\bibitem{ren2019repeatnet}
{\sc P.~Ren, Z.~Chen, J.~Li, Z.~Ren, J.~Ma, and M.~de~Rijke}, {\em Repeatnet: A
  repeat aware neural recommendation machine for session-based recommendation},
  in AAAI, 2019.

\bibitem{rendle2012bpr}
{\sc S.~Rendle, C.~Freudenthaler, Z.~Gantner, and L.~Schmidt-Thieme}, {\em Bpr:
  Bayesian personalized ranking from implicit feedback}, in UAI, 2009.

\bibitem{rendle2010factorizing}
{\sc S.~Rendle, C.~Freudenthaler, and L.~Schmidt-Thieme}, {\em Factorizing
  personalized markov chains for next-basket recommendation}, in WWW, 2010.

\bibitem{sarwar2001item}
{\sc B.~Sarwar, G.~Karypis, J.~Konstan, and J.~Riedl}, {\em Item-based
  collaborative filtering recommendation algorithms}, in WWW, 2001.

\bibitem{tan2016improved}
{\sc Y.~K. Tan, X.~Xu, and Y.~Liu}, {\em Improved recurrent neural networks for
  session-based recommendations}, in Proceedings of the 1st Workshop on Deep
  Learning for Recommender Systems, 2016.

\bibitem{vaswani2017attention}
{\sc A.~Vaswani, N.~Shazeer, N.~Parmar, J.~Uszkoreit, L.~Jones, A.~N. Gomez,
  {\L}.~Kaiser, and I.~Polosukhin}, {\em Attention is all you need}, in
  NeurIPS, 2017.

\bibitem{velivckovic2017graph}
{\sc P.~Veli{\v{c}}kovi{\'c}, G.~Cucurull, A.~Casanova, A.~Romero, P.~Lio, and
  Y.~Bengio}, {\em Graph attention networks}, 2018.

\bibitem{wang2019recurrent}
{\sc J.~Wang and J.~Caverlee}, {\em Recurrent recommendation with local
  coherence}, in WSDM, 2019.

\bibitem{wang2020next}
{\sc J.~Wang, K.~Ding, L.~Hong, H.~Liu, and J.~Caverlee}, {\em Next-item
  recommendation with sequential hypergraphs}, in SIGIR, 2020.

\bibitem{wang2020time}
{\sc J.~Wang, R.~Louca, D.~Hu, C.~Cellier, J.~Caverlee, and L.~Hong}, {\em Time
  to shop for valentine's day: Shopping occasions and sequential recommendation
  in e-commerce}, in WSDM, 2020.

\bibitem{wang2019neural}
{\sc X.~Wang, X.~He, M.~Wang, F.~Feng, and T.-S. Chua}, {\em Neural graph
  collaborative filtering}, in SIGIR, 2019.

\bibitem{wu2019session}
{\sc S.~Wu, Y.~Tang, Y.~Zhu, L.~Wang, X.~Xie, and T.~Tan}, {\em Session-based
  recommendation with graph neural networks}, in AAAI, 2019.

\bibitem{zheng2018spectral}
{\sc L.~Zheng, C.-T. Lu, F.~Jiang, J.~Zhang, and P.~S. Yu}, {\em Spectral
  collaborative filtering}, in RecSys, 2018.

\bibitem{zhu2019improving}
{\sc Z.~Zhu, J.~Wang, and J.~Caverlee}, {\em Improving top-k recommendation via
  jointcollaborative autoencoders}, in WWW, 2019.

\end{thebibliography}

\end{document}